\documentclass[pdflatex,sn-nature]{sn-jnl}
\usepackage{natbib}
\usepackage{geometry}
\usepackage{graphicx}%
\usepackage{multirow}%
\usepackage{amsmath,amssymb,amsfonts}%
\usepackage{amsthm}%
\usepackage{mathrsfs}%
\usepackage[title]{appendix}%
\usepackage{xcolor}%
\usepackage{textcomp}%
\usepackage{manyfoot}%
\usepackage{booktabs}%
\usepackage{algorithm}%
\usepackage{algorithmicx}%
\usepackage{algpseudocode}%
\usepackage{listings}%
\usepackage{rotating} 
\usepackage{booktabs}
\usepackage{tabularx}
\usepackage{multirow}
\usepackage{fvextra}
\usepackage{array}
\DefineVerbatimEnvironment{promptbox}{Verbatim}{fontsize=\footnotesize,breaklines=true,breakanywhere=true}
\usepackage{siunitx}
\usepackage{makecell}
\usepackage{enumitem} 
\usepackage{soul}    
\usepackage{rotating}
\usepackage{url}   %

\theoremstyle{thmstyleone}%

\theoremstyle{thmstyletwo}%
\theoremstyle{thmstylethree}%
\begin{document}

\title[Article Title]{Politicized Attention Shifts Amplify Polarization in the Information Ecosystem during California Wildfires}

\author[1]{\fnm{Yiheng} \sur{Chen}}
\equalcont{These authors contributed equally to this work.}

\author[2]{\fnm{Alina} \sur{Hagen}}
\equalcont{These authors contributed equally to this work.}

\author[3]{\fnm{Fan} \sur{Yang}}

\author[4]{\fnm{Ratna B.} \sur{Dougherty}}

\author[5]{\fnm{Zihui} \sur{Ma}}

\author*[2]{\fnm{Lingyao} \sur{Li}}\email{lingyaol@usf.edu}

\author*[1]{\fnm{Runlong} \sur{Yu}}\email{ryu5@ua.edu}

\affil[1]{\orgdiv{Department of Computer Science},
\orgname{University of Alabama},
\orgaddress{\city{Tuscaloosa}, \state{AL}, \postcode{35487}, \country{USA}}}

\affil[2]{\orgdiv{School of Information},
\orgname{University of South Florida},
\orgaddress{\city{Tampa}, \state{FL}, \postcode{33620}, \country{USA}}}

\affil[3]{\orgdiv{School of Journalism and Mass Communications},
\orgname{University of South Carolina},
\orgaddress{\city{Columbia}, \state{SC}, \postcode{29208}, \country{USA}}}

\affil[4]{\orgdiv{School of Public Affairs},
\orgname{University of South Florida},
\orgaddress{\city{Tampa}, \state{FL}, \postcode{33620}, \country{USA}}}

\affil[5]{\orgdiv{Department of Aerospace Engineering},
\orgname{University of Maryland},
\orgaddress{\city{College Park}, \state{MD}, \postcode{20742}, \country{USA}}}

\abstract{Wildfires require governments to communicate under conditions of urgency, uncertainty, and intense public scrutiny, yet such communication now unfolds within a digitally mediated environment shaped by polarization and engagement-based amplification. We analyze over 1.3 million wildfire-related social media posts from California (2016–2025) to examine how institutional actors are evaluated within this landscape. Users' stance toward government is actor-specific: individual political officials are discussed more negatively than operational agencies across federal, state, and local levels, and this gap widens during extreme wildfire events. Moreover, interaction networks become increasingly modular over time, consolidating into polarized communities in which negativity concentrates within cohesive clusters. Engagement-weighted measures show that highly interactive negative content disproportionately shapes visible discourse, while crisis periods redirect attention from emergency agencies to high-profile political figures, reinforcing reputational divergence. These findings indicate that wildfire communication operates within a polarized, engagement-ranked ecosystem in which evaluative tone, network structure, and visibility dynamics jointly shape institutional perception. Effective disaster communication should therefore account for the structural conditions of contemporary digital public communities.}

\keywords{California Wildfire Analysis, Government Trust, Social Network Analysis, Information Ecosystem, Political Polarization, User Behavior Dynamics}

\maketitle

\section{Introduction}
\label{sec1}

Wildfires represent one of the most severe and rapidly intensifying natural disasters in the United States, with successive fire seasons in California displacing communities, straining infrastructure through mass evacuations and unprecedented economic losses~\cite{kelley2025state}, and intensifying public scrutiny of government response efforts~\cite{mergel2013framework}. During such events, timely and credible communication from official actors is essential for minimizing loss of life and enabling effective emergency response~\cite{FEMAframework, calfireEvacGuide}. Yet, we find that disaster communication no longer unfolds within a stable informational environment; instead, it occurs within a complex information ecosystem shaped by digital media platforms, algorithmic visibility, and user engagement dynamics. Information does not simply transmit but flows through interconnected structures that jointly shape what becomes visible and salient to the public~\cite{benkler2018network_ch2}.

An information ecosystem is a dynamic system of interconnected actors and technologies that produce and circulate information, with human, technical, and social components that must be understood in relation to one another \cite{nstc2022informationintegrity}. Within this ecosystem, the identity of the speaker strongly shapes how messages are interpreted and evaluated \cite{ross2024shortcuts, lodge2005automaticity}. Citizens do not evaluate ``government'' as a unitary entity \cite{brady2022fifty}; rather, they respond to specific institutional actors, including elected officials, administrative agencies, and emergency responders. This distinction is especially important amid long-term declines in public trust in government \cite{brady2022fifty}, as institutional confidence is not evenly distributed across actors. Political leaders and operational agencies occupy distinct reputational positions and are evaluated through different lenses: individual political figures are more likely to be filtered through partisan and affective orientations, whereas agencies and expert communicators are more commonly assessed in terms of competence, expertise, and procedural reliability \cite{schwaderer2025whom}. Treating ``government'' as a single, indivisible entity may therefore obscure meaningful differences in how public stance is directed and how disaster communications are received.

Meanwhile, the digital platforms that structure this ecosystem provide unprecedented observational capacity \cite{dong2021review}. Over recent decades, disaster communication has migrated onto social media~\cite{farnham2005disaster,redcross2012socialmedia,purohit2025engage}, where crisis discourse unfolds at scale and in real time. Unlike surveys or opinion polls, social media captures spontaneous reactions to unfolding events~\cite{dong2021review, li2021data, ma2025analyzing}, enabling longitudinal analysis of stance and discourse as crises evolve and compound. Prior research has demonstrated the value of these data for disaster monitoring and public opinion tracking, while also identifying methodological challenges, including irrelevant content, semantic ambiguity, and difficulty extracting actor-specific perceptions~\cite{li2025mining}. 

However, these digital traces do not simply reveal public opinion \cite{robertson2024inside}; they are generated within platform architectures that favor attention and engagement, which can unevenly distribute visibility \cite{twitterGithub, cinelli2021echo, kitchens2020understanding}. Highly central actors can redirect collective attention during acute events, shaping information flows both within their immediate network communities and across clustered segments of the broader discourse \cite{hunt2024horizontal, cinelli2021echo, kitchens2020understanding}. The question is therefore not merely whether social media reflects public opinion, but how the digital information ecosystem structures the perception, amplification, and evaluation of institutional actors during crises. Although prior studies have identified the benefits of social media as an effective communication tool to share and collect critical data during and after crises~\cite{fraustino2017social, sari20242021, saroj2020use}, far less is known about how amplification mechanisms and evolving network structures interact to redistribute users' attention across different categories of institutional actors under conditions of heightened risk.

In this study, we address these challenges by conducting a large-scale, longitudinal analysis of wildfire-related discourse in California using over 1.3 million social media posts collected between 2016 and 2025. We employ an entity-level extraction approach to identify references to specific government actors and assess online users' stance toward those actors over time. By integrating computational methods with established interdisciplinary scholarship, we address the following questions:

\begin{itemize}
    \item \textbf{RQ1. Government evaluation is heterogeneous across actors and contexts:}
    How does public evaluation of \emph{government} wildfire actors vary across administrative levels and representation types (individual vs.\ organizational), and how do these differences manifest across counties and during severe-fire periods?

    \item \textbf{RQ2. Diffusion reshapes what becomes \emph{visible}:}
    To what extent do repost-driven diffusion and platform-mediated amplification systematically tilt the \emph{visible} tone of government-related wildfire discourse toward a highly engaging minority, and are the resulting shifts concentrated around discrete, event-linked shocks?

    \item \textbf{RQ3. Community structure constrains exchange and reroutes crisis attention:}
    How does the interaction network structure of wildfire discourse evolve over time, and during crises does attention become increasingly routed through segmented communities and a small set of elite hubs, thereby changing which actor classes dominate \emph{negative visibility}?
\end{itemize}


We find that online users' evaluation is systematically actor-specific. Individual political officials are discussed more negatively than organizational entities across federal, state, and local levels, and this gap widens during identified severe wildfire events  \cite{deadlyCALFIREreport,destructiveCALFIREreport, largeCALFIREreport}. Engagement-weighted measures show that highly interactive negative content disproportionately shapes visible discourse, particularly when framed around political conflict. Meanwhile, interaction networks become increasingly modular after 2023, with cross-community bridging declining and negativity concentrating within cohesive clusters. Crisis periods not only increase discourse volume; they also redistribute attention toward elite actors and alter how stance diffuses. These findings indicate that wildfire communication operates within a polarized, engagement-ranked information ecosystem in which institutional evaluation is shaped by both human attribution processes and platform-mediated visibility dynamics. Understanding this ecosystem is essential for designing disaster communication strategies that remain effective under network segmentation and algorithmic amplification.

\section{Methods}
\label{sec:methods}

We study the California wildfire communication \emph{information ecosystem} by combining (i) entity-targeted institutional stance extraction from social media posts, (ii) visibility-weighted stance measures that separate expression from reach, and (iii) interaction-network polarization diagnostics. These components map to institutional heterogeneity (Fig.~\ref{fig:fig1}), amplification and event-linked shocks (Fig.~\ref{fig:weighted_metrics} and Fig.~\ref{fig:amplification}), and structural polarization (Fig.~\ref{fig:network_evolution}).

\subsection{Data source and preprocessing}
We retrieve public posts from X (formerly Twitter) via the Brandwatch API~\cite{Brandwatch2026} (field list and query execution details in Supplementary S1). Posts published between January 1, 2016 and February 12, 2025 are collected using the query
\[
(\texttt{"wildfire"} \;\texttt{OR}\; \texttt{"bushfire"} \;\texttt{OR}\; \texttt{"California fire"} \;\texttt{OR}\; \texttt{"CAfire"} \;\texttt{OR}\; \texttt{"CALfire"})
\]
Posts were retrieved via the Brandwatch API and restricted to California using Brandwatch-inferred \texttt{Region} metadata (definitions and limitations: Supplementary S1), yielding a raw corpus of $N=1{,}349{,}990$ posts. We then applied a standardized preprocessing workflow—text normalization, removal of empty-content records, exact-text deduplication to reduce structural redundancy, and conservative bot filtering using a curated account list (workflow and rules: Supplementary S2). On the preprocessed corpus, we ran a closed-world LLM annotation pipeline to extract eligible institutional entities and assign per-entity stance, producing 1{,}120{,}831 entity instances from 670{,}488 posts; 730{,}337 instances could be linked to counties and were retained for county-level analyses.

\subsection{Entity-level stance annotation with LLMs}

We build an LLM annotation pipeline to (i) extract entities mentioned in each post and (ii) assign \emph{entity-specific} stance labels (\textit{positive}, \textit{negative}, \textit{neutral}) based on explicit evaluative intent. \cite{ALDayel2021StanceSurvey} Entities are restricted to three categories central to institutional wildfire discourse: \textit{government}, \textit{non-profit}, and \textit{news}. Category boundaries are hard-coded to minimize label drift: \textit{government} covers public agencies, offices/officials, and political parties \cite{usc2000cc5}; \textit{non-profit} includes NGOs, charities/foundations, advocacy groups, coalitions, research and educational/cultural institutions, and non-profit media \cite{boris1998nonprofit,usc501c3}; and \textit{news} is limited to for-profit mass media outlets and journalists affiliated with for-profit media \cite{usc2_1602_11}. Because short posts often mention multiple targets, entity-level stance attribution reduces target-misattribution relative to post-level stance aggregation.

To support corpus-scale annotation, the model is required to return a strict JSON array that is programmatically parsed and validated~\cite{Dagdelen2024StructuredIE} (prompt contract, parsing, and failure handling in Supplementary S4; full prompt in Supplementary S11). We select \texttt{gpt-5-mini} for corpus-scale annotation based on performance on an adjudicated $N=200$ human-annotated set (protocol and metrics in Supplementary S5; summary in Table S2). When a post mentioned multiple eligible entities, we retain one row per post-entity instance; repeated labels for the same post-entity pair are collapsed by majority vote using fixed tie-breaking rules (Supplementary S2).

\subsection{Stance aggregation and statistical models}
We map stance labels to numeric scores $s\in\{-1,0,+1\}$ (negative $=-1$, neutral $=0$, positive $=+1$). For any aggregation unit $g$, let $N_g^{+}$, $N_g^{-}$, and $N_g^{0}$ denote the numbers of positive, negative, and neutral labels, and let $V_g = N_g^{+}+N_g^{-}+N_g^{0}$ denote the entity-mention volume. We summarize stance using the polarity rate
\[
PR_g=\frac{N_g^{+}-N_g^{-}}{V_g}.
\]

\textit{Aggregation units.} For Fig.~\ref{fig:fig1}a, we aggregate to county-year $(c,y)$ and compute $PR_{c,y}$; county-years with low support are hatched (threshold and mapping rules in Supplementary S3). For Fig.~\ref{fig:fig1}b, we compute category-year polarity as the unweighted mean of county-year polarity within each category. For Fig.~\ref{fig:fig1}c-d, we aggregate to government entity-day $(e,t)$ and compute $PR_{e,t}$.

\textit{Mixed-effects models.} To quantify within-government divergence by administrative level and representation, we fit linear mixed-effects models~\cite{Bates2015lme4} with entity-specific random intercepts. Let $y_{e,t}=PR_{e,t}$, $L_e\in\{\text{local},\text{state},\text{federal}\}$ denote government level, and $R_e\in\{0,1\}$ denote representation (0=individual, 1=organizational). We estimate
\[
y_{e,t}=\beta_0+\alpha_{L_e}+\gamma R_e+\delta_{L_e}R_e+b_e+\varepsilon_{e,t}.
\]
To test context dependence during statewide severe-fire windows, let $FW_t$ be a day-level indicator defined from consolidated acute-fire intervals (Supplementary Table S8; construction in S3). We estimate
\[
y_{e,t}=\beta_0+\alpha_{L_e}+\gamma R_e+\delta_{L_e}R_e+\kappa FW_t+\lambda(R_e\times FW_t)+b_e+\varepsilon_{e,t}.
\]
We report estimated marginal means (EMMs)~\cite{emmeans2025} with 95\% CIs and planned organizational-minus-individual contrasts (Supplementary Table S7).

\subsection{Visibility-weighted amplification metrics and event analysis}
To separate expressed stance from visibility, we construct user-weighted, engagement-weighted, and impression-weighted stance series on repost streams. Let $\mathcal{P}_t$ be the set of posts in period $t$ and $\mathcal{U}_t$ the set of active users in $t$. Let $\ell_i\in\{\text{neg},\text{neu},\text{pos}\}$ denote the stance label of post $i$ in the repost stream. For negativity, define $s_i=\mathbb{I}(\ell_i=\text{neg})$. Let $e_i$ denote engagement, defined as likes$+$reposts$+$replies, and let $m_i$ denote impressions (views). Here $t$ indexes weeks after weekly aggregation of reposts (Fig.~\ref{fig:weighted_metrics}a--b).

The user-weighted negative rate is
\[
R^{\text{usr}}_t=\frac{1}{|\mathcal{U}_t|}\sum_{u\in\mathcal{U}_t}\frac{\sum_{i\in\mathcal{P}_{u,t}} s_i}{|\mathcal{P}_{u,t}|},
\]
where $\mathcal{P}_{u,t}\subseteq\mathcal{P}_t$ denotes posts authored by user $u$ in period $t$. Attention-weighted negative rates are
\[
R^{\text{eng}}_t=\frac{\sum_{i\in\mathcal{P}_t} s_i\,e_i}{\sum_{i\in\mathcal{P}_t} e_i},\qquad
R^{\text{imp}}_t=\frac{\sum_{i\in\mathcal{P}_t} s_i\,m_i}{\sum_{i\in\mathcal{P}_t} m_i}.
\]
We define user-baseline interaction amplification $\Delta_{E-U,t}=R^{\text{eng}}_t-R^{\text{usr}}_t$ and user-baseline exposure amplification $\Delta_{I-U,t}=R^{\text{imp}}_t-R^{\text{usr}}_t$. We additionally report the engagement--exposure gap $\Delta_{E-I,t}=R^{\text{eng}}_t-R^{\text{imp}}_t$. Analogous definitions apply for positivity (Fig.~\ref{fig:weighted_metrics}b). Field availability and smoothing choices are reported in Supplementary S7. Peak detection, event labeling, and interrupted time-series specifications are detailed in Supplementary S8 and S9~\cite{Bernal2017ITS}.

\subsection{Network construction and polarization metrics}
We construct annual user interaction networks from reply and reshare/repost ties. Nodes represent users and directed edges point from the acting user to the referenced account; edge weights count interaction frequency within a year. For comparability and readability across years, we retain the largest connected component and restrict each annual network to the top $2{,}000$ users by degree. Communities are detected using Leiden on a symmetrized undirected weighted projection~\cite{Traag2019Leiden}.

\textit{Weighted modularity.}
For an undirected weighted graph with edge weights $w_{ij}$, node strength $k_i=\sum_j w_{ij}$, and total weight $2m=\sum_{ij} w_{ij}$, modularity is
\[
Q=\frac{1}{2m}\sum_{ij}\left(w_{ij}-\frac{k_i k_j}{2m}\right)\mathbb{I}(c_i=c_j),
\]
where $c_i$ denotes community assignment~\cite{Newman2006Modularity}.

\textit{E-I index.}
Let $W_{\mathrm{intra}}=\sum_{i<j:\,c_i=c_j} w_{ij}$ and $W_{\mathrm{inter}}=\sum_{i<j:\,c_i\neq c_j} w_{ij}$ denote total within-community and between-community edge weight in the undirected projection. The E--I index~\cite{Krackhardt1988EI} is
\[
\mathrm{E\!-\!I}=\frac{W_{\mathrm{inter}}-W_{\mathrm{intra}}}{W_{\mathrm{inter}}+W_{\mathrm{intra}}}.
\]
We additionally report the share of \textit{bridging nodes} and the bimodality coefficient of user-level stances (details in Supplementary S10), and align layouts across years for visual comparability (see in Supplementary S10).

\textit{Bridging nodes and bimodality.} A node is counted as a \textit{bridging node} if it has at least one neighbor assigned to a different community in the undirected projection. To summarize whether user-level stance is unimodal versus polarized, we compute the bimodality coefficient
\[
\mathrm{BC}=\frac{g_1^2+1}{g_2+3},
\]
where $g_1$ is sample skewness and $g_2$ is sample excess kurtosis (so $g_2+3$ is Pearson kurtosis)~\cite{Pfister2013BC}.

\section{Results}
\label{sec2}

We characterize the California wildfire information ecosystem by linking (i) \emph{who} is evaluated and how (institutional heterogeneity), (ii) \emph{what} becomes visible under diffusion and platform-mediated amplification, and (iii) \emph{how} interaction communities structure attention and constrain cross-group exchange during crises. We report results in four figure-aligned sections. Fig.~\ref{fig:fig1} establishes systematic stance divergence across institutional types and, within government, across administrative level and representation. Fig.~\ref{fig:weighted_metrics} shows that repost-based diffusion induces a strong visibility bias consistent with a vocal-minority mechanism and generates event-linked shocks. Fig.~\ref{fig:network_evolution} documents increasing polarization and reduced cross-community bridging after 2023. Finally, Fig.~\ref{fig:amplification} shows that crises reroute repost amplification toward shifting elite hubs, changing which actor classes dominate negative visibility.

\subsection{Spatial and institutional heterogeneity in institutional stance}

\begin{figure}
    \centering
    \includegraphics[width=1\linewidth]{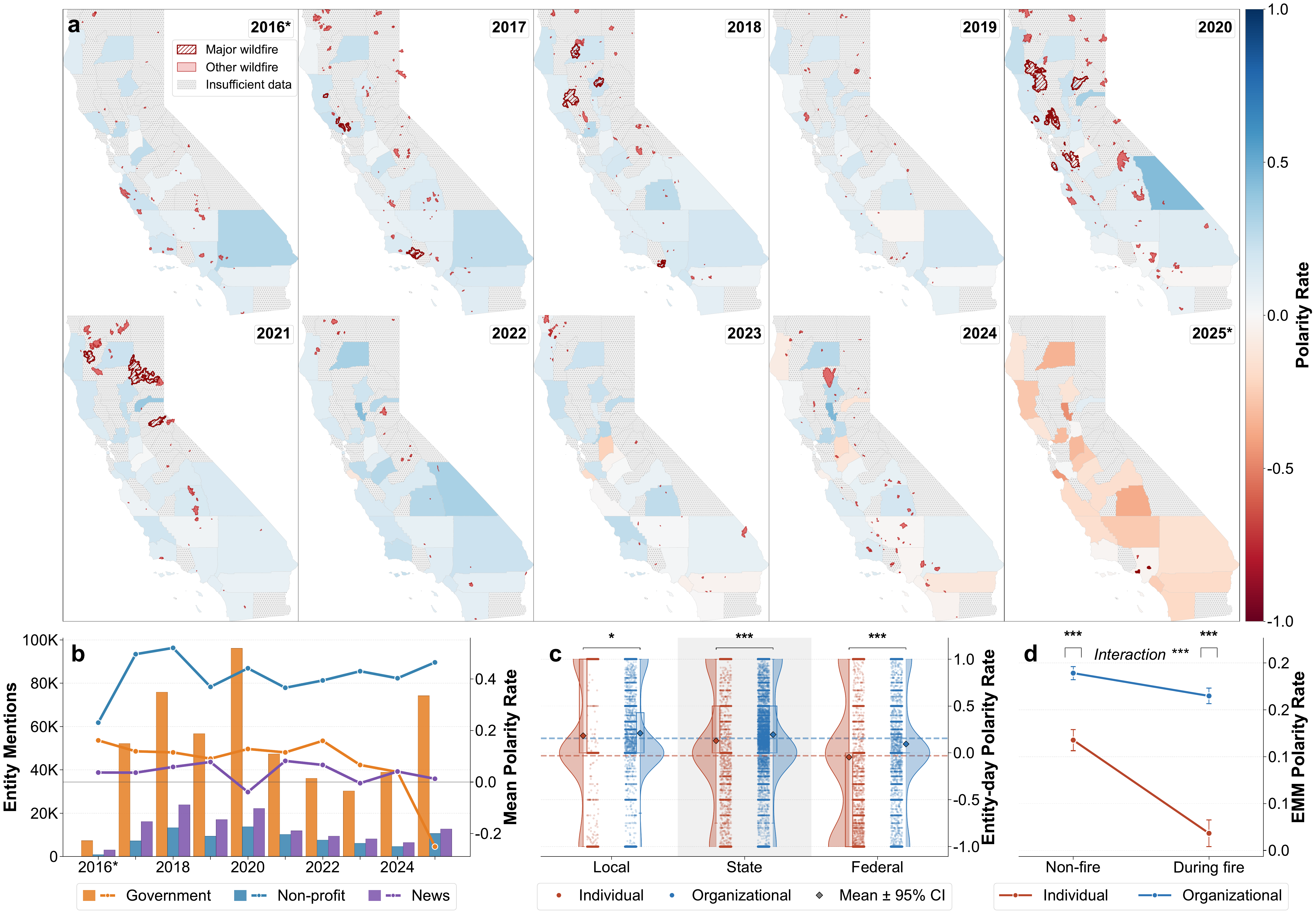}
    \caption{\textbf{Spatial and institutional heterogeneity in institutional stance across California (2016-2025).}
\textbf{a,} County--year polarity rate for counties with at least 30 labeled entity instances; counties below the threshold are hatched. CAL FIRE FRAP wildfire perimeters ($\geq$1{,}000 acres) are overlaid for contextual reference, with major named fires highlighted (see in Supplementary Table S1; S3).
Asterisks indicate years with partial coverage (2016 and 2025), due to incomplete data for the full year.
\textbf{b,} Annual entity-mention volume (bars) and county-mean polarity (lines; unweighted county mean) by entity category.
\textbf{c,} Government entity--day polarity by administrative level and representation (raincloud distributions). Diamonds indicate mixed-effects estimated marginal means (EMMs) with 95\% CIs (see in Supplementary Table S7).
\textbf{d,} Representation-gap context dependence: EMM polarity for organizational vs.\ individual government entities in non-fire versus severe-fire windows (see in Supplementary Table S8).}
    \label{fig:fig1}
\end{figure}

County-year polarity is spatially heterogeneous and varied over time (Fig.~\ref{fig:fig1}a). Polarity is generally positive in 2016--2022 (annual mean $0.1087$--$0.1784$), decline in 2023--2024 (means $0.0914$ and $0.0671$), and is negative in the observed 2025 partial-year window (mean $-0.1913$; median $-0.1773$). County polarity is not monotone in perimeter-defined fire activity: 2020 has the largest total mapped burned area in our $\ge$1{,}000-ac perimeter set (4.14 million acres), yet county polarity remains positive (mean $0.1372$). Fire periods coincide with higher entity-mention volume (Fig.~\ref{fig:fig1}b), whereas the direction of evaluation varies across years and contexts (Fig.~\ref{fig:fig1}a).

Despite spatial and temporal variation, stance follows a consistent ordering by entity type (Fig.~\ref{fig:fig1}b). Government entities dominated mentions in every year (64.66--77.72\%) but exhibited lower county-mean polarity than non-profit organizations, while news remained near neutral. Averaged over 2016--2024, county-mean polarity is $0.1105$ for government, $0.4053$ for non-profit organizations, and $0.0395$ for news; in 2025$\ast$, government polarity drops to $-0.2505$ while non-profit organizations remains strongly positive ($0.4640$). Given the dominance of government actors within the entity set (Supplementary S11), we center subsequent analyses on government-focused discourse.

Within government, stance further diverges by administrative level and representation (Fig.~\ref{fig:fig1}c). A mixed-effects model fit to 151{,}073 entity-day observations across 27{,}384 unique entities detects significant effects of government level, representation, and their interaction (Supplementary Table S7a). Model-based means show that federal individuals are most negative (EMM $=-0.0457$), whereas local and state individuals are positive (EMMs $=0.1830$ and $0.1291$). Organizational actors are more positive than individuals at every level, with the largest organizational advantage at the federal level ($\Delta_{\mathrm{Org-Ind}}=0.1404$), followed by state ($\Delta=0.0663$) and local ($\Delta=0.0267$) (Supplementary Table S7a).

The representation gap also depends on acute wildfire periods (Fig.~\ref{fig:fig1}d; severe-fire windows derived from Supplementary Table~S1, summarized in Table S8). In an entity-day mixed-effects model including an intervention indicator (during severe-fire windows vs.\ non-fire), we detect a significant representation $\times$ intervention interaction (Table S7b). Polarity for individuals declines sharply during fires ($0.1177 \rightarrow 0.0183$), whereas organizational entities remain comparatively stable ($0.1891 \rightarrow 0.1648$); accordingly, the organizational advantage widens from $\Delta=0.0714$ in non-fire periods to $\Delta=0.1465$ during fires (Supplementary Table S7b).

Together, Fig.~\ref{fig:fig1} establishes a stable \emph{supply-side} asymmetry in institutional evaluation. However, these estimates treat posts as equally visible. We next test whether repost diffusion and platform-mediated amplification overweight a highly engaging minority, thereby shifting the stance most \emph{visible} to audiences.

\subsection{Visibility-weighted stance and vocal-minority amplification in government reposts}

\begin{figure}
    \centering
    \includegraphics[width=1\linewidth]{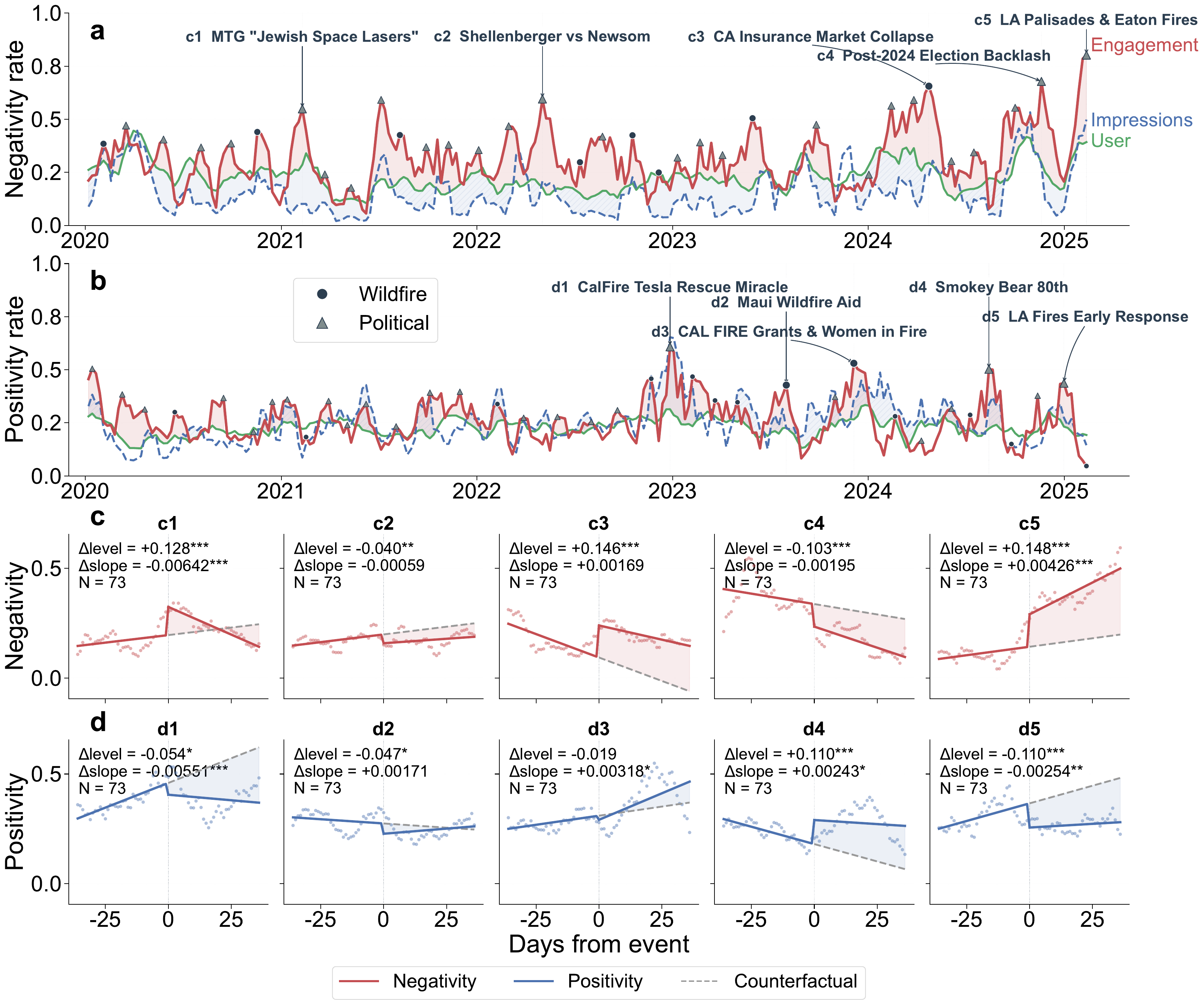}
    \caption{\textbf{Visibility-weighted stance reveals vocal-minority amplification and event-linked shocks in government wildfire discourse.}
\textbf{a,} Weekly negativity rate among posts mentioning government entities under three measurement channels: user-weighted baseline, engagement-weighted salience (likes$+$reposts$+$replies), and impression-weighted exposure (views). Curves show a 4-week rolling mean (2020 onward; smoothing and field availability in S7). Markers (c1-c5) indicate the five largest negative-shock events identified from the smoothed series (event detection and labeling in Supplementary S8; full list in Table Supplementary S5).
\textbf{b,} Weekly positivity rate under the same three channels and smoothing; labeled events (d1-d5).
\textbf{c,} Interrupted time-series (ITS) fits for the negativity events (c1-c5) using daily engagement-weighted negativity within event windows. Solid lines show fitted segments; dashed lines show counterfactual trends under no interruption (Supplementary S9).
\textbf{d,} Analogous ITS fits for positivity events (d1-d5) on daily engagement-weighted positivity (Supplementary S9).}
    \label{fig:weighted_metrics}
\end{figure}

To isolate diffusion rather than authorship, we restrict this analysis to reposts that explicitly mention government agencies (daily series, 2020-01-01--2025-02-12). Across this repost stream, stance estimates depend strongly on the weighting scheme: engagement-weighted negativity is consistently higher and more volatile than the user-weighted baseline (Fig.~\ref{fig:weighted_metrics}a). This pattern is consistent with a vocal-minority mechanism in which a small subset of highly engaging reposts disproportionately shapes observed negative tone.

The divergence is most pronounced at peak moments. At the largest negative surge (LA Palisades \& Eaton Fires), engagement-weighted negativity reaches 0.802 compared with 0.392 under user-weighting ($\Delta_{E-U}=+0.410$) and 0.498 under impression-weighting ($\Delta_{E-I}=+0.304$). Similar amplification gaps appear across the top negative peaks (e.g., $\Delta_{E-U}$ ranging from +0.263 to +0.417; Fig.~\ref{fig:weighted_metrics}a). For positive peaks, the engagement--impression relationship is weaker and can reverse (e.g., ``CalFire Tesla Rescue Miracle'': $\Delta_{E-U}=+0.294$ but $\Delta_{E-I}=-0.046$; Fig.~\ref{fig:weighted_metrics}b), indicating that exposure (``what reaches many users") and interaction (``what provokes response") are separable mechanisms.

Using a BERTopic-assisted~\cite{grootendorst2022bertopic} peak inventory (\(N=44\) peaks per series), we find that negative engagement surges are most often triggered by politicized narratives. Coding peak labels into coarse families shows that political/partisan or governance-conflict framing accounts for a majority of negative peaks (\(n=25\)), with the remainder distributed across economic-system disruptions (e.g., utilities/insurance/market stress), mis-/disinformation framings, and incident-anchored events (Supplementary Table S5; Supplementary S8). In contrast, positive peaks concentrate around preparedness/mitigation and operational response/aid themes, with policy support and commemoration recurring as secondary motifs.

Event-centered interrupted time-series analyses indicate that these peaks correspond to discrete shifts rather than transient noise (Fig.~\ref{fig:weighted_metrics}c--d; Supplementary S9). Politicized controversy produces sharp but rapidly decaying negative shocks (c1: \(\Delta\)level\(=+0.13\), \(p<0.001\); \(\Delta\)slope\(=-0.006\), \(p<0.001\)), whereas the 2025 ``LA Palisades \& Eaton Fires" peak exhibits both an immediate jump and sustained post-event growth (c5: \(\Delta\)level\(=+0.15\), \(p<0.001\); \(\Delta\)slope\(=+0.004\), \(p<0.001\)). Together, Fig.~\ref{fig:weighted_metrics} shows that engagement-based amplification can systematically tilt government-mention repost discourse toward a highly engaging minority, especially under politicized framing.

Visibility bias can be amplified or constrained by interaction structure. We next test whether wildfire discourse communities became more polarized over time, potentially shaping how crisis attention flows across groups.

\subsection{Interaction structure and rising polarization in wildfire discourse}

\begin{figure*}[t]
    \centering
    \includegraphics[width=\linewidth]{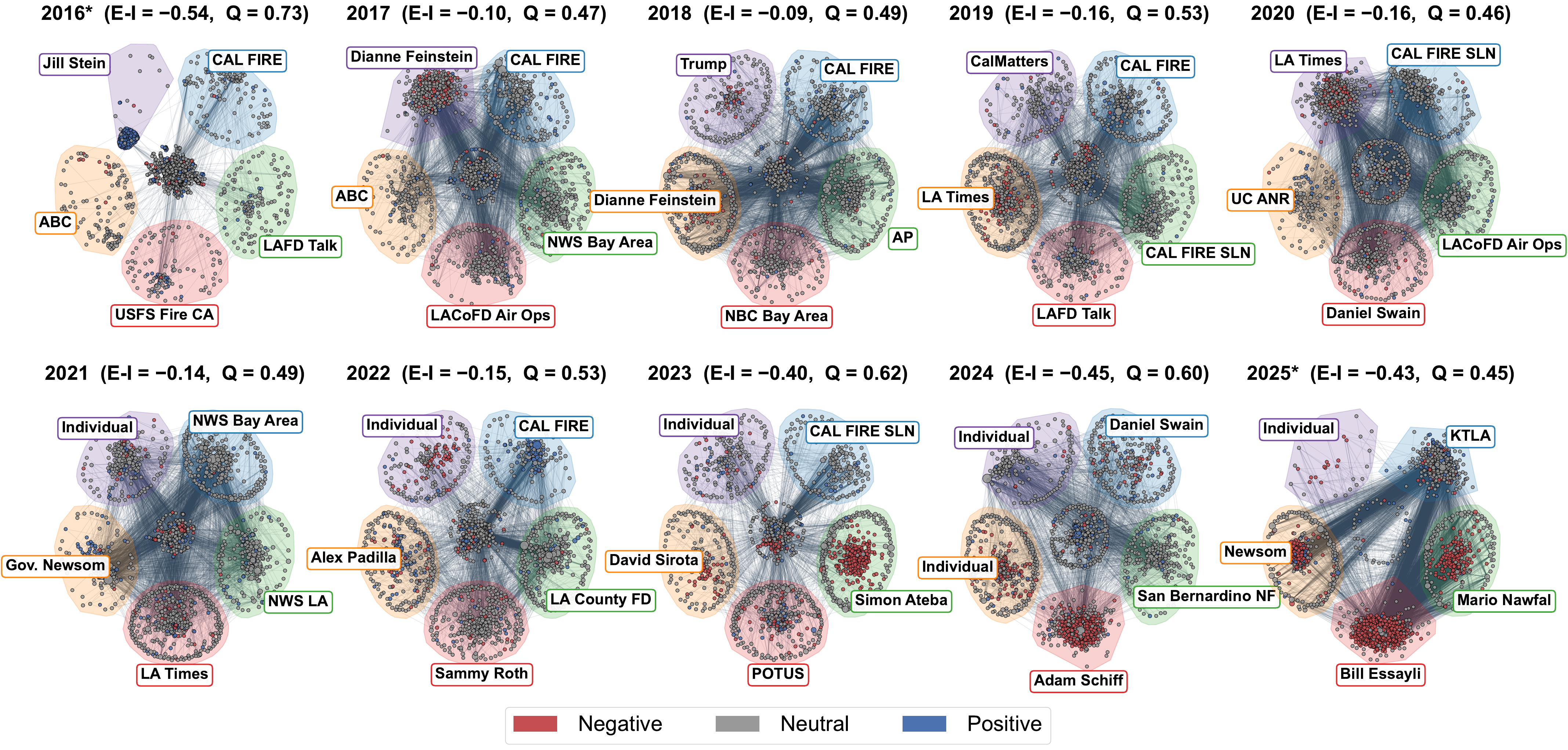}
    \caption{\textbf{Interaction networks show increasing polarization in California wildfire discourse (2016--2025).}
Annual user interaction networks were constructed from reply/repost/share ties. Nodes represent users, and edge weights reflect interaction frequency within a year. Each network is restricted to the largest connected component and the top 2{,}000 users by degree for comparability and readability (Supplementary S10). Communities were detected using Leiden on an undirected weighted projection. Node colour indicates user-level stance (majority stance across posts) and node size scales with degree; only inter-community edges are shown for clarity. Panel titles report weighted modularity $Q$ and the E--I index. ``Individual'' refers to non-institutional users, i.e., accounts not affiliated with institutional media or government entities. Asterisks indicate years with partial coverage (2016 and 2025), due to incomplete data for the full year.}
    \label{fig:network_evolution}
\end{figure*}

Annual interaction networks exhibit persistent community structure rather than a single well-mixed discussion arena (Fig.~\ref{fig:network_evolution}). Across all years, weighted modularity remains high ($Q=0.45$--$0.73$), indicating stable community segmentation in wildfire-related interactions.

From 2017--2022, segregation is comparatively moderate: $Q$ stays near $\sim 0.46$--$0.53$ and the E--I index remains close to zero (approximately $-0.09$ to $-0.16$), consistent with substantial cross-community contact. In this period, bridging nodes are common ($\sim 75\%$--$86\%$ of users), and stance distributions are weakly polarized (BC $\sim 0.12$--$0.19$).

A sharp inflection emerges in 2023. Modularity increases to $Q\approx 0.62$ (2023) and remains elevated in 2024 ($Q\approx 0.60$), while the E--I index becomes substantially more negative (from $\approx -0.15$ in 2022 to $\approx -0.40$ in 2023 and $\approx -0.45$ in 2024). Cross-community bridges thin in parallel: the bridging-node share declines from $\sim 75\%$ (2022) to $\sim 66\%$ (2023) and $\sim 59\%$ (2024). Stance polarization intensifies as well, with BC rising to $\approx 0.33$ (2023) and $\approx 0.36$ (2024), and remaining high in partial-year 2025 (BC $\approx 0.54$).

This structural hardening coincides with a shift in which accounts anchor major communities. Earlier networks feature institutional and operational information hubs (e.g., weather and fire-response accounts) alongside mainstream outlets, whereas post-2023 networks are increasingly centered on individual political figures and partisan influencers, alongside a recurring ``Individual'' cluster (Fig.~\ref{fig:network_evolution}). Consistent with this shift, negativity becomes concentrated within cohesive factions: among the five largest communities, the most negative community increases from $\sim 27.9\%$ negative-majority users (2022) to $\sim 61.3\%$ (2023) and $\sim 64.5\%$ (2024), reaching $\sim 78.1\%$ in partial-year 2025.

Given this increasingly polarized interaction structure, we next ask how major fire crises restructure the \emph{diffusion} layer itself---that is, which elite hubs and actor classes dominate repost amplification and, in particular, negative visibility.

\begin{figure*}[ht]
    \centering
    \includegraphics[width=\linewidth]{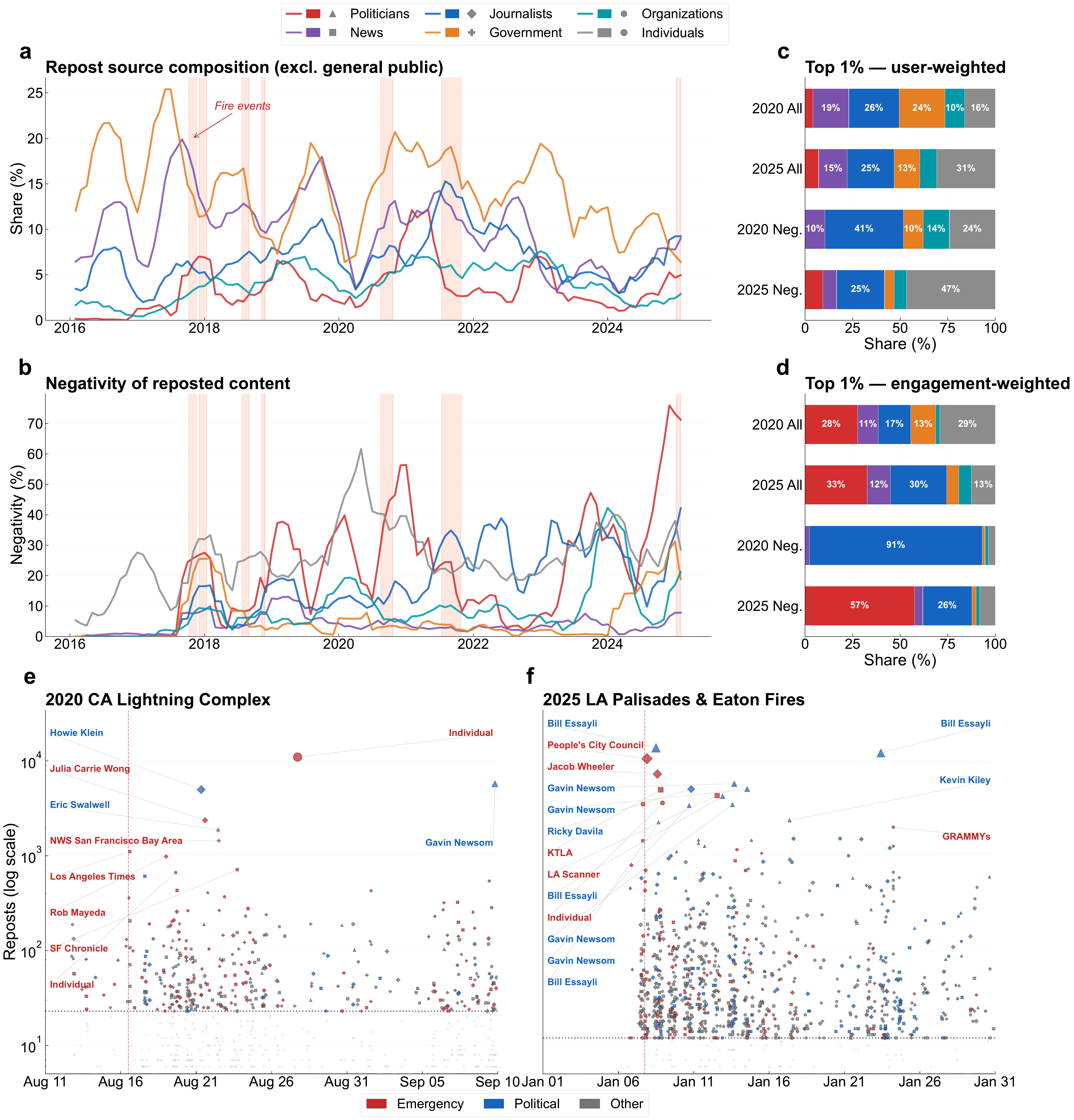}
    \caption{\textbf{Crises reroute repost amplification and shift which actor classes drive negative visibility.}
\textbf{a,} Monthly composition of reposted \emph{sources} by author class (smoothed; sparse months excluded), where sources are the original authors of reposted content (Supplementary S6; smoothing in Supplementary S7).
\textbf{b,} Negativity rate of reposted content by source class over time (same smoothing as in a).
\textbf{c,} Composition of the top 1\% most-amplified source accounts during the 2020 California Lightning Complex window (Aug.~11--Sep.~10) and the 2025 Los Angeles fires window (Jan.~1--31), shown for all reposted content versus negative-only reposted content (user-weighted: each top account contributes equally).
\textbf{d,} The same decomposition under engagement-weighting, emphasizing which source classes account for the largest share of interaction with reposted content.
\textbf{e,f,} Viral-post timelines for each crisis window, showing repost counts for high-velocity posts and annotating dominant themes and source classes (definitions and thresholds in Supplementary S7).}
    \label{fig:amplification}
\end{figure*}

\subsection{Crisis-time rerouting of repost amplification and negative visibility}

Crisis windows repeatedly reconfigure who is amplified and what tone is amplified (Fig.~\ref{fig:amplification}a--b). The repost ecosystem is not evenly distributed across actor types: during major fires (shaded bands), institutional sources (government agencies, news organizations, and journalists) periodically take larger shares of reposted content, indicating that crises change which actor classes become central in diffusion rather than merely increasing activity volume.

Tone amplification is similarly source-dependent. Negativity in reposted content diverges sharply by account type and exhibits event-linked surges (Fig.~\ref{fig:amplification}b). Reposted content attributed to politicians displays the most pronounced spikes in negativity during crisis windows, whereas reposted content from government agencies remains comparatively lower and more stable; journalists and news media occupy an intermediate regime, with negativity rising in acute periods but less persistently than political accounts. These patterns imply that crisis-time diffusion can concentrate negative visibility in actor classes that are not necessarily the most frequently evaluated in the underlying corpus (Fig.~\ref{fig:fig1}b).

To isolate elite amplification in the diffusion channel, we define opinion leaders as the top 1\% most-amplified source accounts (by repost volume) and decompose their reposts under user- versus engagement-weighted aggregation (Fig.~\ref{fig:amplification}c--d). Under the user-weighted view, the top 1\% remains institutionally mixed in both crisis years (Fig.~\ref{fig:amplification}c). Engagement-weighting reveals stronger concentration: attention shifts toward politicians and journalists in 2025 (together $\sim$60\%) relative to 2020 (Fig.~\ref{fig:amplification}d). Negative repost engagement shows the sharpest divergence: in 2020, journalists account for $\sim$90\% of negative engagement within the top 1\%, whereas in 2025 politicians account for $\sim$57\% (Fig.~\ref{fig:amplification}d), indicating distinct crisis pathways through which negative stance becomes visible.

Event-level case studies further show heavy-tailed repost amplification and identifiable elite hubs (Fig.~\ref{fig:amplification}e--f). During the 2020 California Lightning Complex window (Fig.~\ref{fig:amplification}e; $p_{99}=23$ reposts), the extreme tail is led by a first-person eyewitness account (10{,}964 reposts) and a post from Governor Gavin Newsom (5{,}728), with emergency-information and legacy-media accounts forming the next tier. During the 2025 LA Palisades \& Eaton fires (Fig.~\ref{fig:amplification}f; $p_{99}=12$ reposts), the extreme tail is more clearly anchored in political amplification, led by repeated high-repost posts from Bill Essayli (13{,}739 and 12{,}100 reposts), with additional political/commentary accounts in the 5{,}000--10{,}000 range and local news/scanner-style accounts around 4{,}000--5{,}000 reposts. Together with the post-2023 structural hardening (Fig.~\ref{fig:network_evolution}), Fig.~\ref{fig:amplification} shows that crises reroute attention through a small set of opinion leaders, and that the identity of those leaders conditions whether negative visibility is media-led (2020) or politician-led (2025).

\section{Discussion}\label{sec12}

Across a decade of California wildfire discourse on X, we show that government evaluation during wildfires operates through a three-layer crisis information ecosystem: actor-specific attribution, engagement-ranked amplification, and polarized routing. Consequently, the tone that becomes most \emph{visible} during crises can diverge from the tone most \emph{common}, and crisis attention is increasingly rerouted toward elite political hubs.

A key implication of the representation gap is that \textbf{accountability in crisis discourse is personalized}: identifiable officials are more exposed to affect-driven blame than operational agencies. The persistence of this gap across administrative levels---most pronounced for federal actors---suggests a stable accountability pattern that may be intensified when political identity and responsibility attribution become salient, rather than being driven by any single incident. Prior work suggests that elected officials and personal accounts often attract disproportionate engagement relative to administrative organizations \cite{hagen2018government, mickoleit2014social, swalve2025centralized}, positioning identifiable actors as high-visibility nodes when attention becomes politicized; in this setting, personalization can entail reputational vulnerability as officials become efficient targets for negative visibility during crisis peaks. 

These patterns align with psychological mechanisms through which crises intensify evaluation of individuals: political actors are particularly exposed to rapid, affect-laden evaluations \cite{slovic2007affect, lodge2005automaticity}, affective impressions can bias subsequent interpretation of messages \cite{taber2006motivated, stenzel2025negative}, named individuals elicit stronger reactions than diffuse organizations \cite{jenni1997explaining, smart2005devil}, and political psychology distinguishes trust in institutions from evaluations of officeholders \cite{schwaderer2025whom, levi2000political}. Attribution theory further suggests that, under negative conditions, responsibility may be assigned to identifiable actors even when causes are structural \cite{weiner2013human}, aligning with politicized accountability narratives; for disaster communication, the implication is not that personalization is uniformly harmful, but that it can increase reach while simultaneously increasing exposure to affect-driven (and often politicized) evaluation during acute events.

Our visibility-weighted measures speak to work on \textbf{algorithmic amplification and emotionally valenced diffusion}: what appears salient on the interface can systematically diverge from underlying expression, implying that expression, interaction, and exposure are not interchangeable measures of public evaluation and that highly interactive content can disproportionately define what audiences perceive---a pattern consistent with vocal-minority amplification. A plausible mechanism lies in engagement-ranked visibility architectures, where predicted interaction can shape what is repeatedly surfaced rather than chronology or institutional authority \cite{twitterGithub}; emotionally valenced content---especially negative or outrage-linked content---can diffuse more readily because it increases interaction probability \cite{vosoughi2018spread, cao2025does, de2021sadness}. 

Consistent with this logic, our peak inventory indicates that \textbf{engagement and exposure surges are not confined to hazard conditions alone}: peaks often align with governance conflict and politicized narratives, and exposure spikes can coincide with exogenous political and policy moments, suggesting that wildfire discourse is politically permeable and can experience discrete, event-linked shocks in visible tone (Fig.~\ref{fig:weighted_metrics}c--d). These dynamics create a structural tension for government disaster communication: established guidance emphasizes clarity, timeliness, and actionable information to reduce uncertainty and support protective action \cite{reynolds2005crisis, mileti1990communication, fema2026chemicalkpf3, sutton2014warning, hughes2012evolving}, yet ranked visibility can favor content that provokes reaction; while moderate emotional tone may enhance comprehension \cite{megalakaki2019effects}, highly intense framing yields mixed effects on information integration \cite{cao2025does, arfe2023effects, lang2007cognition, jimenez2012emotional}, pointing to a potential misalignment between communication optimized for protective action and the interaction signals that drive visibility under politicized crisis attention.

Beyond actor-specific evaluation and ranked amplification, \textbf{interaction structure conditions where crisis attention can travel}. As wildfire discourse becomes more segmented, cross-community exposure and correction weaken, increasing the likelihood that politicized interpretations are locally reinforced rather than broadly contested \cite{kitchens2020understanding, liu2025conceptualizing}; engagement-ranked systems can further intensify this pattern by prioritizing interaction-efficient content that aligns with learned preferences \cite{gillespie2022not}. Combined with homophilic network formation \cite{bahns2017similarity} and conformity dynamics \cite{ENNSpiralofSilence}, moderate or cross-cutting perspectives may become less visible even when privately held, and clustered diffusion with selective expression can produce the \emph{appearance} of local consensus and apparent extremity, even when ecosystem-wide agreement is limited. 

In polarized environments on X, \textbf{neutral positions can be structurally disadvantaged in diffusion \cite{cinelli2021echo}, and perceived local majorities can discourage dissent within clusters} \cite{ENNSpiralofSilence}; although such self-moderation is difficult to observe directly, the combination of segmented interaction networks and engagement-ranked amplification can narrow the visible range of viewpoints and make highly visible criticism appear more representative than it is, encouraging reactive recalibration that overweights politicized segments while missing quieter operational needs. These structural dynamics intersect with crisis-time rerouting of diffusion through elite hubs: post-2023, central positions are increasingly occupied by identifiable individuals rather than organizational accounts (Fig.~\ref{fig:network_evolution}); organizational communication is typically guided by procedural norms \cite{fema2026chemicalkpf3, reynolds2005crisis}, whereas individual political actors may employ more personalized and affectively expressive frames \cite{enli2013personalized, dang2017investigation} that are interaction-efficient in ranked systems. When such actors occupy central positions in modular networks, their interpretations can diffuse disproportionately \cite{hunt2024horizontal}, reinforcing politicized attention loops; consistent with this logic, crises reconfigure which actor classes dominate amplified repost attention and, in particular, negative visibility (Fig.~\ref{fig:amplification}), highlighting that polarization not only segments audiences but can also reshape who serves as an attention gatekeeper during crises.

These findings suggest three practical implications for government disaster communication in engagement-ranked and polarized environments. First, although personalized communication can increase engagement \cite{hagen2018government, mickoleit2014social}, identifiable officials appear especially vulnerable to affect-driven blame during crises; organizational accounts may therefore function as stabilizing channels for procedural guidance when politicized attention spikes, and response strategies that emphasize concern, corrective action, and consistency may reduce reputational spirals when responsibility is contested \cite{kwok2022crisis}. Second, engagement metrics should not be treated as direct proxies for public evaluation: highly interactive negativity can dominate visibility even when not representative of broader expression, motivating multi-channel monitoring of expression, engagement, and exposure when evaluating reception. Third, in increasingly modular networks, reaching across communities may require partnerships with trusted intermediaries and cross-cutting hubs \cite{hunt2024horizontal}, because politicized clusters can otherwise monopolize visible crisis interpretation and limit the diffusion of operational guidance; collaborating with locally trusted actors may help ensure that evacuation orders, mitigation updates, and safety information reach vulnerable or institutionally skeptical communities that might otherwise remain outside dominant diffusion pathways.

Several limitations qualify the interpretation of these findings. First, we analyze a single platform (X), and observed amplification and polarization dynamics may reflect platform-specific affordances that do not generalize uniformly across digital environments. Second, exposure-based measures rely on impression fields that are consistently available only in later years, which constrains long-run comparability across visibility channels. Third, social media discourse constitutes a behaviorally consequential information environment but is not a representative survey of population attitudes \cite{robertson2024inside}. Fourth, although entity-targeted stance extraction reduces misattribution relative to post-level sentiment, it relies on automated LLM classification within a predefined entity scope, which may introduce systematic labeling error or exclude relevant actors. 

Future work could strengthen causal identification and deepen interpretation of the mechanisms suggested by our results through cross-platform comparisons, qualitative validation or systematic content analysis of framing and rhetoric (especially contrasting political individuals versus operational agencies), and quasi-experimental designs leveraging exogenous shocks (e.g., sudden fire onsets, policy announcements, or platform-level changes). Beyond operational strategy, these patterns imply that visible opinion climates during wildfires are shaped by politicized attention allocation and polarized diffusion, not simply by aggregated attitudes: high-engagement signals and repeated within-community exposure can act as social cues about what others believe or endorse \cite{udry2024illusory}, potentially shaping interpretation and behavioral intention \cite{hale2002theory}. In this setting, treating visible discourse as a direct proxy for public evaluation risks over-weighting highly amplified signals while underestimating how segmentation constrains the reach of operational guidance; as social platforms remain central infrastructures for emergency information, policymakers and communicators should account for how politicized attention shifts, amplification incentives, and community segmentation jointly condition perception during wildfire crises.

\backmatter




\section*{Declarations}

\begin{itemize}
\item Competing interests. \\ 
The authors declare no competing interests.
\item Data availability. \\
The raw data used in this study were collected via Brandwatch and cannot be publicly shared due to the platform's data privacy and security policy (\url{https://www.brandwatch.com/legal/}). Researchers seeking access to the original X data must obtain it through the official platform channels and in accordance with applicable policies. For academic and non-commercial purposes regarding data access procedures, the X post IDs associated with the dataset will be made available by the corresponding authors upon request, allowing for independent verification and replication. The processed dataset used for analysis in this study, including derived variables, annotations, and aggregated measures, is publicly available at \url{https://doi.org/10.5281/zenodo.18820058}.
\item Code availability. \\
The code files are publicly available at: \url{https://github.com/defene/Information-Ecosystem}.
\item Author contributions. \\
F.Y. and L.L. collected the data. 
Y.C. and A.H. preprocessed the data and contributed to study design, interpretation of results, and manuscript writing. 
Y.C. developed the analytical code. 
F.Y., R.B.D., and Z.M. contributed to data analysis, interpretation of results, and revision of the manuscript. 
L.L. and R.Y. conceived the study, supervised the project, and contributed to study design, interpretation of results, and revision of the manuscript. 
Y.C. and A.H. contributed equally to this work. 
All authors reviewed and approved the final manuscript.

\end{itemize}

\clearpage

\bibliography{references-revised-yc}

\end{document}